%% file: 2021-rce-clustering-workflow-graphs.tex
\documentclass[conference,hidelinks]{IEEEtran}
\IEEEoverridecommandlockouts
\usepackage{amssymb}
\usepackage{graphicx}
\usepackage{fullpage}
\usepackage{xspace}
\usepackage{amsmath}
\usepackage{booktabs}
\usepackage{tikz}
\usetikzlibrary{shapes, calc}

\usepackage{cite}
\usepackage{hyperref}
\usepackage{cleveref}

\usepackage[inline]{enumitem}

\input{preamble}

\title{Towards Automated Semantic Grouping in Workflows for Multi-Disciplinary Analysis\thanks{This work is based on the Bachelor Thesis of the first author~\cite{Schneider2020ClusteringRCEWorkflow}.}}

\author{\IEEEauthorblockN{Dominik Schneider}
\IEEEauthorblockA{\textit{Institute for Software Technology} \\
\textit{German Aerospace Center (DLR)}\\
51147 \Cologne, Germany \\
dominik.schneider@dlr.de}
\and
\IEEEauthorblockN{Alexander Weinert}
\IEEEauthorblockA{\textit{Institute for Software Technology} \\
\textit{German Aerospace Center (DLR)}\\
51147 \Cologne, Germany \\
alexander.weinert@dlr.de}
}

\begin{document}

\maketitle

\begin{abstract}
  When designing multidisciplinary tool workflows in visual development environments, researchers and engineers often combine simulation tools which serve a functional purpose and helper tools that merely ensure technical compatibility by, e.g., converting between file formats.
  If the development environment does not offer native support for such groups of tools, maintainability of the developed workflow quickly deteriorates with an increase in complexity.

  We present an approach towards automatically identifying such groups of closely related tools in multidisciplinary workflows implemented in \RCE by transforming the workflow into a graph and applying graph clustering algorithms to it.
  Further, we implement this approach and evaluate multiple clustering algorithms.
  Our results strongly indicate that this approach can yield groups of closely related tools in \RCE workflows, but also that solutions to this problem will have to be tailor-made to each specific style of workflow design.
\end{abstract}


\section{Introduction}
\label{sec:introduction}

In multidisciplinary research projects, engineers and scientists from multiple disciplines collaborate to analyze complex technical or social systems with respect to a multitude of properties.
Such systems include, e.g., airplanes, energy systems, or traffic systems.
Each discipline contributes software tools which simulate individual phenomena or calculate individual KPIs.
Workflow designers then compose these tools into a simulation workflow using visual development environments.

A typical workflow consists of relatively few ``major'' tools, which simulate technical or social phenomena, and numerous ``helper'' tools.
These helper tools, e.g., pre- or post-process data, convert between discipline-specific file formats, or ensure for data integrity.
During an engineering project the number of major and helper tools as well as the complexity of the connections between them grows due to increases in fidelity of modeling and analysis.

Visual development environments usually do not, however, indicate the strong cohesion between groups of major tools and their helper tools.
In spite of this, workflow designers must maintain the relation between major tools and their associated helper tools to ensure correctness of the workflow.
Thus, iterating the design of workflows in the visual design is cumbersome and error-prone.

We show an approach to identifying such clusters of semantically related tools in workflows implemented in Remote Component Environment (\RCE).
This approach uses only static knowledge about the relation between the tools and does not leverage information about a data model underlying the investigated system.
To this end, we transform workflows into directed weighted graphs to which we apply graph clustering methods.
We evaluate our approach on real-world workflows and discuss how the principles used here may generalize to other graphical descriptions of complex tool chains.

\paragraph*{Structure of this Work}

After presenting the visual development environment \RCE and basic nomenclature in Section~\ref{sec:background}, we define the problem of finding groups of closely related tools in Section~\ref{sec:tool-groups}.
We then present our approach for determining such groups in engineering workflows defined in \RCE in Section~\ref{sec:framework}.
We show an example of how we instantiate that approach for and evaluation and apply that instantiation to three example workflows in Section~\ref{sec:example}.
Finally, we discuss current limitations and future work in Section~\ref{sec:conclusion}.

\subsection*{Related Work}
\label{sec:related-work}

A complete overview over all works on graph clustering is out of the scope of this work.
We instead refer to the survey paper by Schaeffer~\cite{Schaeffer2007GraphClustering} for a comprehensive overview.
In Section~\ref{sec:example} we discuss clustering algorithms based on properties of potential clusters~\cite{Newman2004FindingEvaluatingCommunity} and on the edges connecting potential clusters~\cite{Newman2003MixingPatternsCommunity}.
Algorithms for computing either measure have been refined and improved since their initial publication~\cite{Donetti2004DetectingNetworkCommunities,Du2007AlgorithmDetectingCommunity,Danon2006EffectSizeHeterogeneity,Yoon2006AlgorithmModularityAnalysis,Dunn2005UseEdgeBetweenness,Brandes2001FasterAlgorithmBetweenness}.

Graph clustering has been applied successfully in numerous fields, e.g., in the analysis of biological processes~\cite{Fortunato2010CommunityDetectionGraphs,Junker2011AnalysisBiologicalNetworks,Yoon2006AlgorithmModularityAnalysis}, social networks~\cite{Fortunato2010CommunityDetectionGraphs,Wasserman1994SocialNetworkAnalysis,Newman2001StructureScientificCollaboration}, or business processes~\cite{Tanaka2012WorkflowSchedulingMinimize,Jung2009HierarchicalClusteringBusiness}.

Furthermore, instead of finding structures within a graph, clustering approaches have also been used to determine sets of similar graph structures~\cite{Santos2008FirstStudyClustering,Jung2006WorkflowClusteringMethod}.

Previous work on determining similar software relies on, e.g., data on users interested in the software~\cite{ZhangLoKochharEtAl2017,NguyenDiRoccoRubeiEtAl2018}, the source code of the software~\cite{KawaguchiGargMatsushitaEtAl2006,McMillanGrechanikPoshyvanyk2012,NguyenDiRoccoRubeiEtAl2018}, or meta-data on the software~\cite{ZhangLoKochharEtAl2017,NguyenDiRoccoRubeiEtAl2018}.
In contrast, our approach only leverages the data-flow between software when applied to solve a particular problem.

Numerous graphical environments other than \RCE allow engineers and scientists to develop, maintain, and execute multi-disciplinary analyses.
One such environment is \ModelCenter by \PhoenixIntegration~\cite{ModelCenter}.
In contrast to \RCE, neither the source-code nor the binary of \ModelCenter is freely available.

There furthermore exist \ApacheNifi~\cite{ApacheNifi} and \Knime~\cite{Knime}, both of which feature a graphical editor for developing tool workflows.
Instead of allowing the integration of black-box tools for performing arbitrary analyses and computations, these tools focus more on performing identical operations on all elements of large data sets.
Thus, their main area of application lies in the field of data science and not in that of multi-disciplinary engineering.

\section{Background and Nomenclature}
\label{sec:background}

In this section, we first describe Remote Component Environment (\RCE) in Section~\ref{sec:background:rce} and give a brief overview over its capabilities for defining and executing workflows.
We furthermore give a brief overview over graphs and graph clustering in Section~\ref{sec:background:clustering}.

\subsection{RCE Workflows}
\label{sec:background:rce}

There are numerous visual environments (cf. Section~\ref{sec:related-work}) that support engineers and scientists with developing and orchestrating multi-disciplinary tool workflows.
These environments offer vastly different capabilities and mechanisms for users to define the data flow between tool instances.
In this work, we focus on workflows as defined in \RCE, an open-source-application developed at DLR.
\RCE allows its users to, among other features, integrate disciplinary tools, to define connections between them via a graphical interface, and to execute the resulting workflow.
In contrast to many similar environments, \RCE does not require users to define a data-model for the model of the investigated system.

Here, we only describe those parts of \RCE that are relevant to our work.
For a comprehensive overview over \RCE and a presentation of its features, please refer to the work by Boden et al.~\cite{Boden2019RceIntegrationEnvironment}.

\begin{figure*} \centering
  \includegraphics[width=\textwidth]{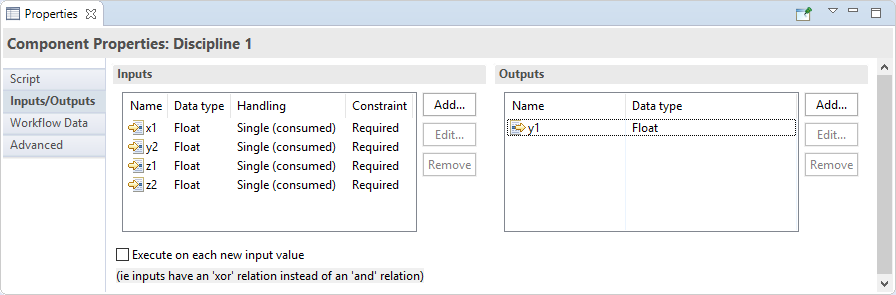}
  \caption{%
    The endpoints of the tool instances ``Discipline 1.'' %
    This instance has five endpoints, four of which are inputs, one of which is an output. %
    All endpoints have a type of Float. %
    We show the context of the tool instance ``Discipline 1'' in Figure~\ref{fig:rce-workflow-editor}. %
  }
  \label{fig:rce-endpoints}
\end{figure*}

When constructing a multi-disciplinary workflow using \RCE, users typically begin by integrating their disciplinary tools into \RCE.
This mainly amounts to defining the \emph{endpoints}, i.e., inputs and outputs, of each tool, scripts that translate endpoint data provided by the API of \RCE into endpoint data for the called tool, and scripts that execute the tool.
During integration, the users specify a \emph{data type} for each endpoint.
At the time of writing, \RCE supports the data types
\begin{itemize*}[label={}]
  \item \booltype,
  \item \inttype,
  \item \floattype,
  \item \vectortype,
  \item \stringtype,
  \item \smalltabletype,
  \item \matrixtype,
  \item \filetype, and
  \item \dirtype.
\end{itemize*}
For inputs, the users furthermore define whether the data arriving at that endpoint are to be treated as an input for a single execution of the tool, or as a parameter to be used in multiple tool executions.
In the former and latter case, we say that an input is \emph{consumed} or \emph{constant}, respectively.
Finally, the users define for each input whether the data arriving at that input is \emph{required} or \emph{optional} for an execution of the tool.
We show the endpoints of a tool in Figure~\ref{fig:rce-endpoints}.

\begin{figure} \centering
  \includegraphics[width=\linewidth]{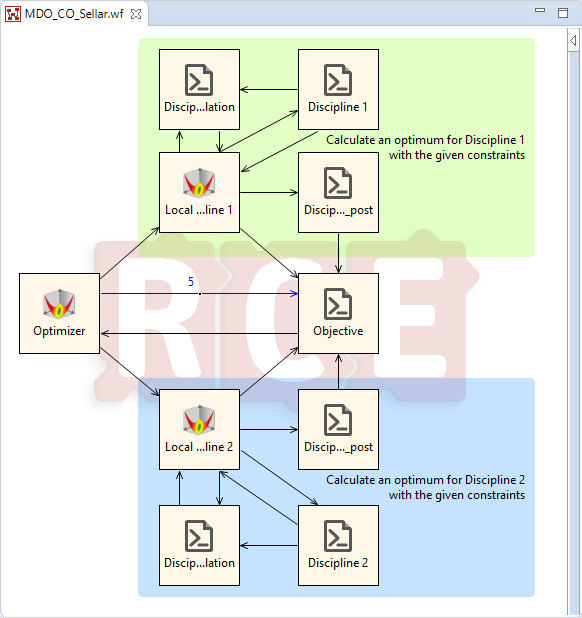}
  \caption{%
    The Workflow Editor of \RCE. %
    The workflow shown contains two tools and ten tool instances. %
    The highlighted edge between the instances ``Optimizer'' and ``Objective'' represents five connections between these instances. %
  }
  \label{fig:rce-workflow-editor}
\end{figure}

Having defined the set of tools required for implementing a multidisciplinary workflow, the users then use a graphical \emph{workflow editor} to place \emph{(tool) instances} on a graphical canvas.
Using this canvas they furthermore graphically define \emph{connections} between tool instances.
We show an example of this graphical editor in Figure~\ref{fig:rce-workflow-editor}.

Each connection connects an output of some instance to an input of some other instance that has the same data type.
A connection has the same data type as the endpoints it connects.
We say that two instances~$s$ and~$t$ are \emph{connected} if there exists a connection going from an output of~$s$ and an input of~$t$.

Besides constructing a workflow, the workflow editor supports the usage of \emph{labels} to mark groups of instances shown in \Cref{fig:rce-workflow-editor} as blue and green rectangles.
Not all instances are assigned to a group in this example.
This is representative of workflows created by practitioners, in which tool instances may belong to multiple groups or none at all.

\subsection{Graphs and Graph Clustering}
\label{sec:background:clustering}

A \emph{graph}~$\graph = (V, E, \weight)$ consists of a finite set of vertices~$V$, a set of edges~$E \subseteq V \times V$, and a weighting function $\weight \colon E \rightarrow \rationals$, where~$\rationals$ denotes the set of positive rational numbers.
If~$(v, v') \in E$, we say that~$v$ and~$v'$ are \emph{adjacent}.
If for each $(v, v') \in E$ we also have $(v', v) \in E$, then we call~$\graph$ \emph{undirected}.
We call the process of constructing an undirected graph from a directed one~$\graph$ \emph{symmetrization} and call the result of that process a \emph{symmetrization of~$\graph$}.

A \emph{clustering of~$\graph$} is a set $\set{\graph_1,\dots,\graph_n}$ of graphs with $\graph_i= (V_i, E_i, \weight_i)$, where all~$V_i$ and all~$E_i$ are pairwise disjoint and that additionally satisfies
\begin{itemize}
  \item $\cup_{1 \leq i \leq n} V_i = V$,
  \item $\cup_{1 \leq i \leq n} E_i \subseteq E$, as well as
  \item $\forall 1 \leq i \leq n, e \in \graph_i.\,\weight_i(e) = \weight(e)$.
\end{itemize}
An \emph{(undirected) clustering algorithm} is an algorithm that takes an (undirected) graph~$\graph$ as input and outputs a clustering of~$\graph$ as output.

\section{Closely Related Tool Instances in Workflows}
\label{sec:tool-groups}

An \RCE workflow comprises multiple tools, each of which may occur in multiple instances in the workflow.
These tools are executed in dependence of one another to simulate, analyze, and optimize numerous facets of the system under investigation.
Most tools functionally depend on other tools in the workflow, e.g., an aerodynamics simulation depends on a structural simulation, which produces a geometric description of some vessel.

In addition to these functional dependencies each tool also has technical requirements and characteristics.
These include, e.g., the file formats produced by the tools or parameter files required for the execution of a tool in the context of the workflow.
Since the tools typically come from different domains with different standards and conventions, these requirements and characteristics are usually heterogeneous.

To construct a fully automated workflow out of these tools with disparate requirements, users usually integrate ``helper tools'' into the workflow.
These helper tools do not contribute directly to the overall calculation of the workflow, but instead, e.g., convert between different data formats, provide parameter files, or perform simple pre- or postprocessing of data.

Thus, completed workflows often feature groups of ``semantically closely related tool instances,'' i.e., tool instances that prepare data for one another and fulfill no functional requirement on their own.
Instead, they collaborate with the other tools in their group to satisfy an overarching computational goal.
Intuitively, these groups of closely related tools are akin to functions in procedural programming.
Similarly to functions, these groups may also be hierarchical, i.e., they may contain other groups of even more closely related tool instances.

Users typically work with these groups of tool instances as if they were a single tool:
If the ``main tool'' is moved to some other location in the graphical editor, the helper tools usually move with them.
If the users remove the main tool from the workflow, the helper tools have to be removed as well.

While the graphical editor of \RCE allows users to move groups of tool instances in unison, it does not feature any further support for editing groups of tool instances simultaneously.
In particular, it displays workflows only in a ``flat'' view, i.e., without allowing users to ``fold away'' or ``group'' parts of a workflow.
This encumbers design and maintainability in large-scale workflows, which can easily grow to contain hundreds of tool instances.
Users have developed stop-gap solutions to improve readability and maintainability of such workflows by, e.g., marking groups of related tool instances with the same background color (cf. Figure~\ref{fig:rce-workflow-editor}).
These serve, however, mainly as visual aids, have to be constructed manually, and are not updated automatically when the visual layout of the workflow changes.

\section{Determining Groups of Closely Related Tool Instances}
\label{sec:framework}

In this section, we present a method that allows users to automatically identify groups of semantically closely related tool instances as described in Section~\ref{sec:tool-groups}.
Since \RCE gives users a great deal of expressive freedom in the design of their workflow, a one-size-fits-all-solution to identifying such groups based solely on the information contained in the constructed workflow is infeasible.
Instead, our method contains customization points that allow the users to adapt this method to their personal preferences of workflow design.

We describe the framework in Section~\ref{sec:framework:overview}.
Subsequently, we describe the customization points in Section~\ref{sec:framework:graph-construction} through Section~\ref{sec:framework:clustering-algorithm}.

\subsection{Framework Overview}
\label{sec:framework:overview}

To automatically determine groups of semantically closely related tool instances in a given workflow, we first transform a workflow~$\wf$ into a graph~$\graph_\wf$ which contains one vertex for each tool instance in~$\wf$.
We then either directly apply a clustering algorithm to that graph, or we construct an undirected graph from it, to which we subsequently apply an undirected clustering algorithm.
In either case, we obtain a clustering $\clustering = \set{\graph_1, \dots, \graph_n}$ of~$\graph_\wf$.

Since the clustering algorithm used to construct~$\clustering$ does not have domain knowledge about the structure of workflows,~$\clustering$ may be too coarse to represent meaningful groups of closely related tool instances in~$\wf$.
Thus, we iteratively apply the clustering algorithm to graphs from~$\clustering$ until the resulting clustering represents meaningful groups of closely related tool instances.

Once~$\clustering$ has the desired resolution, we terminate the iterative application of the clustering algorithm and obtain the desired groups of closely related tools from~$\clustering$.
Since each vertex of each graph in~$\clustering$ directly corresponds to a tool instance in~$\wf$ by construction, this final step is straightforward.

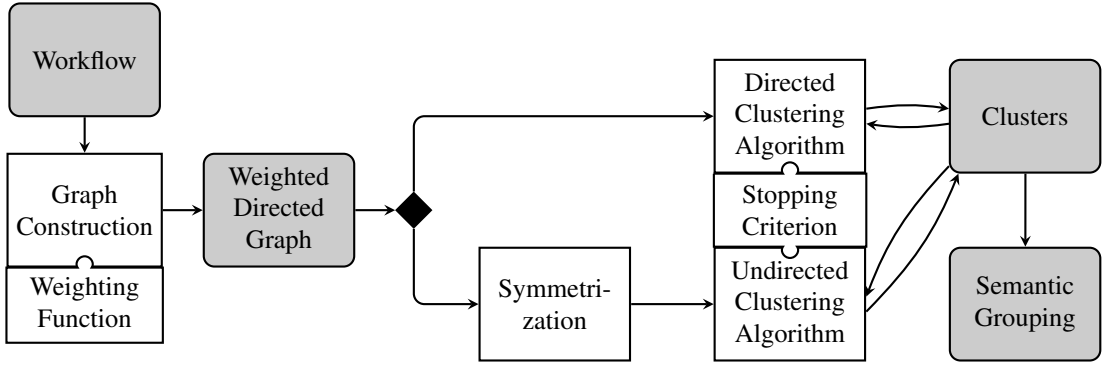
\begin{figure*} \centering
  \input{images/overview}
  \caption{Building-block view of grouping-framework. Gray rectangles indicate objects, while white ones indicate processes. Jigsaw pieces indicate customization points.}
  \label{fig:building-blocks}
\end{figure*}

We illustrate our framework in Figure~\ref{fig:building-blocks}.
In that figure, we furthermore highlight the main customization points of our framework, namely
\begin{itemize*}[label={}]
  \item the weighting function chosen for constructing the graph,
  \item the choice of the clustering algorithm and the stopping criterion deciding when to terminate the iterative application of that algorithm, and
  \item the method of symmetrization, if choosing to use an undirected clustering algorithm.
\end{itemize*}

While for each customization point arbitrary algorithms and criteria can be chosen, each customization point has ``soft'' requirements that contribute to the efficacy of the overall framework.
We identify these requirements in the following sections.

\subsection{Graph Construction}
\label{sec:framework:graph-construction}

The first step in our construction consists of turning a given workflow~$\wf$ into an \emph{associated graph}~$\graph_\wf$.
We define $\graph_\wf = (V, E, \weight)$ with~$V = \set{v_t \mid t\text{ is instance of tool in }\wf}$ and with $(v_s, v_t) \in E$ if and only if an output of instance~$s$ is connected to an input of instance~$t$ in $\wf$.

It remains to define the weighting function~$\weight$.
To this end, we use some insight about the eventual interpretation of~$\graph_\wf$ by a clustering algorithm.
Most clustering algorithms use the weight of an edge $(v, v')$ to determine whether to place~$v$ and~$v'$ in the same cluster.
There is, however, no canonical order on edge weights:
While some clustering algorithms aim to keep vertices that are connected with a high edge weight together in a cluster, others, aim to keep vertices connected with a low edge weight in a cluster.
For example, the agglomerative clustering algorithm, which is one of our applied algorithms described in~\Cref{sec:example:framework-instance}, uses shortest paths to determine similarity and therefore creates clusters with low edge weights within, whereas our application of spectral bisection expresses similarity by high weights~\cite{VonLuxburg2007TutorialSpectralClustering}.

An algorithm that prefers to place vertices connected with high-weight edges in a cluster can, however, easily be transformed into an algorithm that prefers clusters with low-weight edges.
To this end, we replace each weight in a given graph with its reciprocal prior to clustering and revert that operation on each graph of the resulting clustering.
Hence, for the remainder of this work, we assume w.l.o.g. that clustering algorithms interpret high edge weight as a close relation between the two adjacent vertices and aim to retain such edges in the produced clusters.

As discussed in the introduction to this section, \RCE affords its users a large degree of flexibility when constructing workflows.
Thus, it is infeasible to fix a weighting function for associated graphs of arbitrary workflows that results in an effective framework.
Instead, we only formulate the following generic requirement that states that edge weights are well-ordered with respect to the semantic closeness of the tool instances they represent:
\begin{quotation}
There must exist a~$\operatorname{\bowtie} \in \set{<, >}$ such that for each two pairs~$s,t$ and~$s',t'$ of tool instances in~$\wf$, if~$s$ and~$t$ are semantically closer than~$s'$ and~$t'$, then~$\weight(v_s, v_t) \bowtie \weight(v_{s'}, v_{t'})$.
\end{quotation}
We give examples for defining~$\weight(v_s, v_t)$ based on the data exchanged between the tool instances~$s$ and~$t$ in Section~\ref{sec:example:framework-instance}.

\subsection{Symmetrization}
\label{sec:framework:symmetrization}

Some clustering algorithms require their input graphs to be undirected.
Recall, however, that in the previous section we constructed~$\graph_\wf$ as a directed graph:
There exists an edge~$(v_s, v_t)$ in~$\graph_\wf$ if and only if an output of tool instance~$s$ is connected to an input of tool instance~$t$.

Thus, we are unable to apply undirected clustering algorithms to~$\graph_\wf$.
Instead, we first need to symmetrize~$\graph_\wf$, obtaining~$\graph'_\wf$.

A naive symmetrization would intuitively ``drop'' the information on the direction of edges and assign the same weight to both edges between two vertices.
Formally given a graph~$\graph = (V, E, \weight)$, let 
\[ 
  \weight_0(v,v') = \begin{cases}
    \weight(v,v') & \text{if }(v, v') \in E \\
    0             & \text{otherwise.}
  \end{cases}
\] 
We define the \emph{naive symmetrization of~$\graph$} as $(V, E' , \weight')$, where
\begin{itemize}
  \item $E' = \set{(v, v'), (v', v) \mid (v, v') \in E)}$, and
  \item $\weight'(v, v') = \weight_0(v, v') + \weight_0(v', v)$.
\end{itemize}

While this naive approach retains the basic structure of the directed graph, it loses the information encoded in the direction of the edges.
There exist approaches to symmetrization that aim to retain this information.
Here, we highlight bibliometric symmetrization as one such approach~\cite{Satuluri2011SymmetrizationsClusteringDirected}.
Intuitively, bibliometric symmetrization uses the weight of the edges~$(v, v')$ and~$(v', v)$ to denote the number of adjacent vertices shared by~$v$ and~$v'$ as well as the weight of their connections to these vertices.

Recall that the \emph{adjacency matrix} of a graph $(V, E, \weight)$ with $V = \set{v_1,\dots,v_n}$ is the $n\times n$-matrix~$A$ defined as~$a_{ij} = \weight_0(v_i, v_j)$ where~$\weight_0$ is defined as above.
Conversely, for each $n\times n$-Matrix~$A$ over~$\rationals$, there exists a unique graph~$\graph$ (up to isomorphism) such that~$A$ is the adjacency matrix of~$A$.
Further, given a matrix~$A$, we denote the transpose of~$A$ as~$A^T$.

Given the adjacency matrix~$A_\wf$ of~$\graph_\wf$, let~$A'_\wf = A_\wf + I$, where~$I$ denotes the $n\times n$-identity matrix.
We then define the \emph{bibliometric symmetrization of~$\graph_\wf$} as the unique graph with the same vertex set as~$\graph_\wf$ and the adjacency matrix~$(A'_\wf)(A'_\wf)^T + (A'_\wf)^T(A'_\wf)$.

While we highlighted two approaches to symmetrization in this section, this list is not exhaustive:
Other symmetrization techniques are equally valid for use in our framework.
Their efficacy for use in our framework has to be judged on the amount and structure of information they retain in the resulting undirected graph.
It furthermore strongly depends on the construction style of the workflow used as input to the framework.

\subsection{Clustering Algorithm}
\label{sec:framework:clustering-algorithm}

Recall that our definition of a clustering algorithm is very broad and does not place any restrictions on the produced clustering.
We opted for such a broad definition since there is no consensus on how to define the properties of a ``good'' clustering irrespective of the application domain~\cite{Schaeffer2007GraphClustering}.
Further, recall that \RCE affords its users large expressive freedom in constructing their workflow.
Thus, there likely is no single clustering algorithm that recognizes clusters corresponding to meaningful groups of closely related tool instances in arbitrary workflows.

There are, in contrast, some clustering algorithms that are clearly unsuited for use in an effective instantiation of our framework.
More precisely, a clustering algorithm used in our framework should
\begin{enumerate*}[label=\alph*)]
  \item accept a stopping criterion as a parameter and stop the clustering process when the clustering satisfies that criterion, and
  \item should not require any parameters it does not infer from the given input graph.
\end{enumerate*}

We discuss stopping criteria in the following subsection.
The latter requirement precludes the use of clustering algorithms that take, e.g., a fixed number of clusters or the size of clusters to be produced.
Such parameters would either have to be fixed for application to arbitrary workflows or they would have to be inferred from the given workflow prior to application of the clustering algorithm.
In the former case, it would be straightforward to find example workflows that result in which that instantiation of the framework is ineffective.
Inferring the number and size of clusters to be produced from the workflow could, in contrast, greatly improve the efficacy of our framework.
Doing so would, however, require precisely that insight on the structure of arbitrary workflows that we aim to gain using this framework.

\subsection{Stopping Criterion}
\label{sec:framework:stopping-criterion}

Requiring the clustering algorithm to accept a stopping criterion as a parameter allows us to decide whether to terminate clustering the graph or to iterate on the produced clusters.
If we had to rely on criteria hard-coded in the algorithm for this decision instead, we would be unable to determine whether the produced clusters correspond to meaningful groups of closely related tool instances.
This requirement is a rather technical one, as most clustering algorithms used in practice can easily be adapted to terminate after a single ``internal'' iteration.

It remains to determine a stopping criterion for the iterated application of the clustering algorithm.
Again, no single one-size-fits-all criterion is likely to exist due to the expressiveness of \RCE workflows.
For our framework, we follow one approach widely used in literature is:
We pick a metric~$m$ with a bounded value range of~$[0;1]$ and determine a cutoff value~$c$ in that range.
If~$m$ evaluates a graph resulting from a clustering of~$\graph_\wf$ to a value less than~$c$, we apply the clustering algorithm to that graph again.

Widely used metrics include, e.g., cluster density, the global~\cite{Luce1949MethodMatrixAnalysis} and the local~\cite{Watts1998CollectiveDynamics} clustering coefficient, as well as modularity~\cite{Newman2004FindingEvaluatingCommunity}.
The cutoff value~$c$ is again strongly dependent on the design of the actual workflow given as input for the framework.
We report on our experiments with different cutoff values for the metrics given above in the following section.

\section{Example}
\label{sec:example}

In the previous section we have presented our framework for determining groups of semantically closely related tool instances in \RCE workflows.
Due to the expressiveness afforded to users by \RCE, this framework has numerous customization points that allow the users to customize the framework to the workflow they are clustering.
In this section, we evaluate our framework.
To this end, we pick algorithms and criteria for each of the customization points in Section~\ref{sec:example:framework-instance} and apply the instantiated framework to three example workflows described in Section~\ref{sec:example:workflow}.
Finally, we discuss the results of our experiments in Section~\ref{sec:example:discussion}.

\subsection{Instantiating the Framework}
\label{sec:example:framework-instance}

To instantiate the framework, we define
\begin{enumerate*}[label=\alph*)]
\item a weighting function,
\item a clustering algorithm possibly including a symmetrization method, and
\item a stopping criterion.
\end{enumerate*}

We start with defining the weighting function.
As described in Section~\ref{sec:framework:graph-construction}, the main requirement towards the weighting function is to model the semantic closeness between tools.
For our experiments, we use the ``amount of potential data exchange'' between two tool instances as a proxy for this closeness.
We quantify this potential data exchange by analyzing not only the number of connections between two tool instances, but also the type of data transferred via these connections and their ``importance'' for the receiving tool.

Recall that connections between two tool instances in \RCE workflows have
\begin{enumerate*}[label=\alph*)]
\item a data type,
\item a data constraint, and
\item a data handling.
\end{enumerate*}
There may be multiple connections between two instances of tools in a workflow.

Let $\conn_0,\dots,\conn_n$ be the connections between tool instances~$s$ and~$t$ with~$\conn_i = (\datatype_i, \constraint_i, \handling_i)$, where $\datatype_i$, $\constraint_i$, and $\handling_i$ denote the data type, the data constraint, and the data handling of~$\conn_i$, respectively.
Intuitively, we define the weight of the edge~$(v_s, v_t)$ by adding the weight of the transferred data type and a constant offset based on whether the data transferred is obligatory or optional for the execution of~$t$.
Since the data handling is more of a technical property than a semantical one, we omit it in the calculation of the weight of the edge.

Formally, we first define weight functions for data types and constraints in Table~\ref{tab:weighting-functions:datatypes} and Table~\ref{tab:weighting-functions:constraints}.
We choose the weights of data types and constraints arbitrarily based on the intuition that, e.g., a file can transport more information than a float, which can in turn carry more information than a single boolean.

Having defined the weights of data types and constraints, we use these functions to define~$\weight(\datatype, \constraint, \handling) = \weight(\datatype) + \weight(\constraint)$ and obtain the \emph{data-driven weight function} $\weight_d(v_s, v_t) = \Sigma_{0 \leq i \leq n}\weight(\conn_i)$.
To investigate the effect of different orderings on the results of the clustering, we furthermore use the \emph{reciprocal data-driven weight function} $\weight^{-1}_d\colon (v, v') \mapsto (\weight_d(v, v'))^{-1}$ in our experiments.

\begin{table} \centering
  \caption{Weighting Functions for Data Types.}
\label{tab:weighting-functions:datatypes}
\begin{tabular}{lr} \toprule
  Type & $\weight(\text{Type})$ \\ \midrule
  \booltype & 12 \\
  \inttype & 13\\
  \floattype & 14\\
  \vectortype & 15\\
  \dirtype & 16\\ \bottomrule
\end{tabular}\quad
\begin{tabular}{lr} \toprule
  Type & $\weight(\text{Type})$ \\ \midrule
  \stringtype & 17\\
  \smalltabletype & 18\\
  \matrixtype & 19\\
  \filetype & 20\\ \bottomrule \\ 
\end{tabular}
\end{table}

\begin{table} \centering
\caption{Weighting Function for Input Constraints.}
\label{tab:weighting-functions:constraints}
\begin{tabular}{lr} \toprule
  Constraint & $\weight(\text{Constraint})$ \\ \midrule
  None & 0 \\
  Not Required & 3 \\
  Required if Connected & 4 \\
  Required & 5 \\ \bottomrule
\end{tabular}%
\end{table}

To investigate the effect that the weight has on the resulting clustering, we furthermore use the \emph{unit weighting function} $\weight_1\colon (v, v') \mapsto 1$.

Having defined the weighting function, we now describe our choice of clustering algorithms.
There are numerous clustering algorithm, each with their own main areas of application, strengths, and weaknesses.
Our main goal for these experiments is to determine whether ``off-the-shelf'' graph clustering methods are suited for determining groups of closely related tool instances.
Thus, we evaluate
\begin{enumerate*}[label=\alph*)]
  \item edge betweenness clustering~\cite{Newman2003MixingPatternsCommunity},
  \item spectral bisection~\cite{Schaeffer2007GraphClustering,Spielman1996SpectralPartitioningWorks}, and
  \item an adapted version of agglomerative clustering~\cite{Schaeffer2007GraphClustering}
\end{enumerate*}
as clustering algorithms for our framework.
Edge betweenness clustering is based on the idea of removing edges through which most shortest path are running, as these are interpreted as connections between clusters.
Spectral bisection uses the second smallest eigenvalue and the corresponding eigenvector of a graph's Laplacian matrix to split the graph in two clusters.
Agglomerative clustering is a bottom-up approach starting with a single vertex per cluster which are iteratively merged based on a symmetrical distance function.
Here, we use the average path weight of all shortest paths between two clusters as a distance function.
For a more detailed explanation of these algorithms, please refer to the work by Schneider~\cite{Schneider2020ClusteringRCEWorkflow}.

While edge betweenness does not require its input to be an undirected graph, both agglomerative and spectral clustering are undirected clustering algorithms.
Thus, to use them in our framework, we must fix symmetrization techniques.
Here, we use the naive and the bibliometric symmetrization presented in Section~\ref{sec:framework:symmetrization}.
As each clustering algorithm is an undirected clustering algorithm, we also combine both symmetrization approaches with edge betweenness as a clustering algorithm.

Finally, we require a stopping criterion that determines when to terminate the iterated application of the clustering algorithm.
Again, there exist numerous such criteria that measure the ``cohesiveness'' of a given graph.
We combine the chosen algorithms with the stopping criteria presented in Section~\ref{sec:framework:clustering-algorithm}, i.e., with
\begin{enumerate*}[label=\alph*)]
  \item cluster density as defined by Schaeffer~\cite{Schaeffer2007GraphClustering},
  \item the global clustering coefficient~\cite{Luce1949MethodMatrixAnalysis},
  \item the local clustering coefficient~\cite{Watts1998CollectiveDynamics,Kemper2009ValuationNetworkEffects}, and
  \item modularity~\cite{Newman2004FindingEvaluatingCommunity}.
\end{enumerate*}
Cluster density measures the number of edges in a cluster as a fraction of the number of all possible edges in the cluster.
Intuitively, the global clustering coefficient measures the number of fully connected triplets of vertices as a fraction of the number of almost fully connected triplets of vertices.
The local clustering coefficient expresses how close a vertex and its neighborhood is to be a clique and is applicable to a group of vertices by determining the average of each vertex' local clustering coefficient.
Finally, the modularity metric describes the relation between a cluster and the whole graph.
A high modularity indicates the structure of a cluster is more group-like than the graph's structure.
We adapt the standard notion of modularity to range from~$0$ to~$1$ by expressing it as a fraction of the sum of all edge weights in the graph.
For a more thorough explanation of these measures, we again refer to the work by Schneider~\cite{Schneider2020ClusteringRCEWorkflow}.

Each of these criteria determines a score for a graph in the range $[0;1]$.
Thus, we additionally pick a cutoff value and stop the iteration once the metric exceeds that value.
For our experiments, we pick the cutoff values $\set{0.2, 0.4, 0.6, 0.8}$.

\begin{figure*} \centering
  \input{images/example}
  \caption{Instantiations of customization points used in our experiments.}
  \label{fig:instantiations}
\end{figure*}
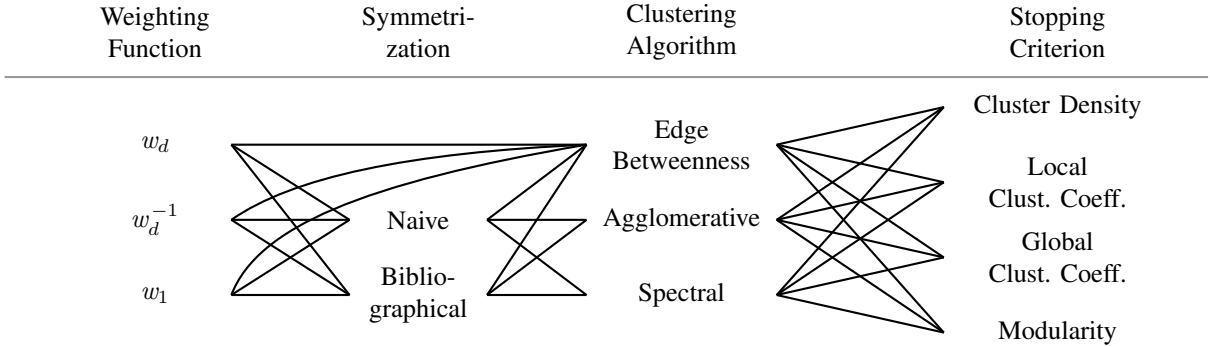

We give an overview over the different values of the customization points in Figure~\ref{fig:instantiations}.

\subsection{Grouping an Example Workflow}
\label{sec:example:workflow}

We evaluate the instantiations of our framework described in the previous section on three real-life \RCE workflows.
The first workflow is included as an example with \RCE, while the second and third workflow were created by users of \RCE for research projects in the field of aerospace engineering.

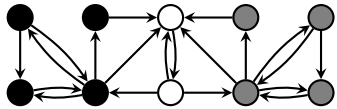
\begin{figure} \centering
  \input{images/MDO_CO_Sellar_labeled}
\caption{Graph representation of Workflow 1 containing three groups, colored in black, white and gray.
The black and gray groups correspond to the blue and green groups shown in Figure~\ref{fig:rce-workflow-editor}, respectively.}
\label{fig:workflow-1-graph}
\end{figure}

We show the first workflow in Figure~\ref{fig:rce-workflow-editor} and give an overview over the basic properties of these workflows in Table~\ref{table:workflow-properties}.
Furthermore, we show the vertices and edges of the associated graph of the first workflow in Figure~\ref{fig:workflow-1-graph}.

\begin{table}\centering
  \caption{Number of Tool Instances, Connections between Tool Instances, and Edges in the Associated Graph of the Evaluated Workflows.}
  \label{table:workflow-properties}
  \begin{tabular}{lrrr} \toprule
  $\wf$ & Tool Instances & Connections & Edges in $\graph_{\wf}$ \\ \midrule
	  Workflow 1 & 10 & 78 & 20\\
	  Workflow 2 & 151& 359 & 194\\
	  Workflow 3 & 262 & 702 & 311\\ \bottomrule
  \end{tabular}
\end{table}

\subsection{Results and Discussion}
\label{sec:example:discussion}

We have implemented the instantiations of our framework for clustering \RCE workflows described above as a stand-alone prototype tool~\cite{Schneider2021}.
This tool parses \RCE workflows, builds their associated graphs using weighting functions based on the weighting of data types and input constraints, and clusters the associated graphs using edge betweenness clustering, spectral clustering, or agglomerative clustering.
We used this tool to evaluate our method on the three real-life workflows described above.
We performed these evaluations on a system with an Intel i7-5700HQ with up to 3.5 GHz per core and 16 GB RAM.
The execution took place within a virtual machine running Linux Ubuntu Server 18.04.5 with eight GB RAM available.

We show the average run time of these executions grouped by the chosen clustering algorithm in Table~\ref{tab:runtimes}.
A run time greater than, say, ten seconds, is clearly prohibitive for interactive usage.
Thus, the results indicate that out of the evaluated clustering algorithms only edge betweenness and spectral clustering are feasible for use in interactive use cases.

\begin{table}\centering
  \caption{Average Runtimes in ms grouped by Clustering Algorithms.}
  \label{tab:runtimes}
  \begin{tabular}{lrrr} \toprule
    Workflow & Edge Betweenness & Agglomerative & Spectral \\ \midrule
    Workflow 1 & 151 & 315 & 587 \\
    Workflow 2 & 2\thinspace690 & 43\thinspace915 & 1\thinspace408 \\
    Workflow 3 & 4\thinspace375 & 260\thinspace441 & -- \\ \bottomrule
  \end{tabular}
\end{table}

Recall that there is no formal definition of a ``good clustering'' or of ``reasonable groups of closely related tool instances.''
Thus, there also is no ground truth against which to evaluate the results of our approach, which makes it impossible to quantitatively evaluate our approach at this point.
While it would be possible to use the background coloring produced by users as a stop-gap solution for the readability of complex workflows (cf. Section~\ref{sec:tool-groups}), these colorings are often incomplete, inconsistent, and strongly dependent on the personal preferences of the users producing them.

Instead we only evaluate the results of the clustering manually.
In general, this manual inspection shows that our experiments produced both clusterings that correspond to reasonable groups of closely related tools as well as clusterings that do not.
As one example, consider the graph representation shown in Figure~\ref{fig:workflow-1-graph}.
Here, our framework inferred multiple clusterings differing by only one vertex from the one given by the users of \RCE.
There is, however, no discernible pattern as to which combinations of customizations yield reasonable results.

Subjectively, the edge betweenness clustering algorithm with the modularity metric as termination condition appears to be the most promising approach. 
Although the modularity metric resulted in usable clusterings across the three workflows, the cutoff score used for achieving these usable clusterings varies.
Besides, different applications of the weighting function do not lead to different results.
Thus, the used input properties for the function are not indicators for a strong relationship between components.

Taking all results into consideration, we conclude that the approach is heavily prone to fluctuations induced by input parameters.
These input parameters could either be determined a priori by a developer or during application by the user.
The former approach would require taking into account a multitude of possible workflow designs.
The latter approach would, in contrast, raise the bar for general applicability of our method and impose further work on the end-user, which might mitigate the increase in productivity gained by this method.

The presence of reasonable clusterings in the results of our experiments indicates, however, that our approach presented in this work is feasible and warrants further investigation.
For a more detailed investigation of the results, we refer to work by Schneider~\cite{Schneider2020ClusteringRCEWorkflow}.

\section{Conclusion}
\label{sec:conclusion}

Our approach shows that a graph representation can easily be derived from an \RCE workflow by using intuitive assumptions and methods.
This enables us to apply widely-known graph analysis techniques with well-defined requirements to loose-defined workflows.
The application of graph clustering methods leads to mixed results.
On the one hand, our experiments were inconclusive with respect to the optimal instantiation of our framework.
On the other hand, we could identify groups within workflows which one might have chosen only by taking the graphical representation into account.

\paragraph*{Future Work}
Recall that our ultimate goal is to support engineers and scientists in developing and maintaining large-scale workflows for analysis and optimization of complex systems.
To quantify this potential improvement, we are planning on performing an empirical study on the gains in productivity and cognitive load obtained by our method.
This empirical study is, however, complicated by issues inherent in measuring the productivity of engineers and scientists~\cite{BrownSvenson1988} as well as by the absence of a wide pool of publicly available engineering workflows.
Further, we expect that the benefits of our work are mainly observed when working with large-scale human-developed workflows.
This precludes the creation of a pool of synthetic examples for the purpose of evaluation.

After obtaining a ground truth of clusterings, we could evaluate our method quantitatively by measuring the ``similarity'' of two clusterings of the same graph~\cite{Tzerpos1999MojoDistanceMetric,ZhihuaWen2003OptimalAlgorithmMojoa,ElRamlyIglinskiStrouliaEtAl2001Modelingsystemuser}.

Currently, we only determine flat clusterings.
It would be straightforward to adapt our framework to yield ``hierarchical'' groups.
The efficacy of such groups for the maintenance would have to be investigated.

In this work we have only used properties of workflows that directly influence the possible executions of the workflow for constructing a graph.
However, workflows constructed in a visual development environment also comprise considerable syntactical features such as, e.g., their absolute and relative positioning.
Such features aid the comprehension of workflows by practitioners, but do not influence the execution of a workflow.
In future work we will investigate the influence of syntactical properties on the identified semantical groups.

Further, we restrict our method to properties that can be inferred from a workflow without executing it.
The run-time-behavior of tools, however, is a black box when only analyzing the workflow statically.
By executing individual tools or complete workflows and inferring information about the relation between inputs and outputs of a tool, it might be possible to determine the semantical relation between tools more precisely.

In this work, we have focused on tool workflows created in \RCE.
It would be interesting to investigate which parts of this framework carry over to other visual workflow development environments.

\bibliographystyle{splncs04}
\bibliography{2021-rce-clustering-workflow-graphs}

\end{document}

%% file: preamble.tex
\newcommand{\set}[1]{\{ #1 \}}

\newcommand{\graph}{G}
\newcommand{\clustering}{\mathcal G}

\newcommand{\rationals}{\mathbb{Q}}

\newcommand{\weight}{\mathit{w}}

\newcommand{\wf}{\mathit{wf}}

\newcommand{\ModelCenter}{Model Center\xspace}
\newcommand{\PhoenixIntegration}{Phoenix Integration\xspace}
\newcommand{\ApacheNifi}{Apache Nifi\xspace}
\newcommand{\Knime}{Knime\xspace}

\newcommand{\RCE}{RCE\xspace}

\newcommand{\conn}{\text{cn}}

\newcommand{\datatype}{\mathit{dt}}
\newcommand{\constraint}{\mathit{cns}}
\newcommand{\handling}{\mathit{hnd}}

\newcommand{\booltype}{\text{bool}}
\newcommand{\dirtype}{\text{dir}}
\newcommand{\filetype}{\text{file}}
\newcommand{\floattype}{\text{float}}
\newcommand{\inttype}{\text{int}}
\newcommand{\matrixtype}{\text{matrix}}
\newcommand{\stringtype}{\text{string}}
\newcommand{\smalltabletype}{\text{smalltable}}
\newcommand{\vectortype}{\text{vector}}

\newcommand{\Cologne}{K{\"o}ln}

%% file: images/overview.tex
  \begin{tikzpicture}[thick,>=stealth,xscale=1.25]
      \tikzstyle{object}=[draw, minimum height=1.5cm, minimum width=2cm, align=center, rounded corners, fill=black!20]
      \tikzstyle{process}=[draw, minimum height=1.5cm, minimum width=2cm, align=center]
      \tikzstyle{decision}=[draw, fill=black, align=center, diamond]

      \node[object]
          (workflow) at (0,2) {Workflow};
      \node[process]
          (construction) at (0, 0) {Graph \\ Construction};
      \node[decision]
          (decision) at (3.5, 0) {};
      \node[object]
          (graph) at ($(construction.east) ! .5 ! (decision.west)$) {Weighted\\Directed\\Graph};

      \node[process]
          (directed-clustering) at (7.5,1.25) {Directed\\Clustering\\Algorithm};

      \node[process]
          (symmetrization) at (5,-1.25) {Symmetri-\\zation};
      \node[process]
          (undirected-clustering) at (7.5,-1.25) {Undirected\\Clustering\\Algorithm};

      \node[object,anchor=north]
        (clusters) at ($(directed-clustering.north) + (2.5,0)$) {Clusters};
      \node[object,anchor=south]
        (groups) at ($(undirected-clustering.south) + (2.5,0)$) {Semantic\\Grouping};

      \newcommand{\puzzle}[2]{
        \coordinate (#1-attach) at (#1.south);
        \draw[fill=white] (#1.south west) -- ($(#1-attach) - (.075,0)$)
            to[out=120,in=180] ($(#1-attach) + (0,.15)$)
            to[out=0,in=60] ($(#1-attach) + (.075,0)$)
            -| ($(#1.south east) - (0,1)$)
            -| cycle;
        \node[anchor=north,align=center] at (#1-attach) {#2};
      }

    \puzzle{construction}{Weighting\\Function}

    \draw[fill=white] (directed-clustering.south west)
            -- ($(directed-clustering.south) - (.075, 0)$)
            to[out=120,in=180] ($(directed-clustering.south) + (0,.15)$)
            to[out=0,in=60] ($(directed-clustering.south) + (.075,0)$)
            -| (undirected-clustering.north east)
            -- ($(undirected-clustering.north) + (.075, 0)$)
            to[out=-60,in=0] ($(undirected-clustering.north) - (0,.15)$)
            to[out=180,in=-120] ($(undirected-clustering.north) - (0.075,0)$)
            -| cycle;

    \node[align=center] at ($(directed-clustering.south) ! .5 ! (undirected-clustering.north)$) {Stopping\\Criterion};
      
    \path[draw]
        (workflow) edge[->,>=stealth] (construction)
        (construction) edge[->,>=stealth] (graph)
        (graph) edge[->,>=stealth] (decision)
        (symmetrization) edge[->,>=stealth] (undirected-clustering)

        ($(directed-clustering.east) + (0,.1cm)$)
          edge[->,>=stealth,bend left=10] ($(clusters.west) + (0,.1cm)$)
        ($(clusters.west) - (0,.1cm)$)
          edge[->,>=stealth,bend left=10] ($(directed-clustering.east) - (0,.1cm)$)

        ($(undirected-clustering.east) + (0,.1cm)$)
          edge[<-,>=stealth,bend left=10] ($(clusters.south west) + (0,.1cm)$)
        ($(clusters.south west) + (.1cm,0)$)
          edge[<-,>=stealth,bend left=10] ($(undirected-clustering.east) - (0,.1cm)$)
        (clusters) edge[->,>=stealth] (groups);

      \path[->,>=stealth,draw,rounded corners]
        (decision) |- (directed-clustering);
      \path[->,>=stealth,draw,rounded corners]
        (decision) |- (symmetrization);
\end{tikzpicture}

%% file: images/example.tex
\begin{tikzpicture}[thick,xscale=4,every node/.style={align=center}]]

  \tikzstyle{weight}=[minimum width=2cm]
  \tikzstyle{symmetrization}=[minimum width=1.8cm]
  \tikzstyle{clustering}=[minimum width=2.5cm]
  \tikzstyle{criterion}=[minimum width=3cm]

  \begin{scope}[shift={(0,.5)}]
    \node[weight] (data-weight) at (0, 0) {$\weight_d$};
    \node[weight] (data-rec-weight) at (0, -1) {$\weight^{-1}_d$};
    \node[weight] (unit-weight) at (0, -2) {$\weight_1$};
  \end{scope}

  \begin{scope}[shift={(0.875,-.5)}]
    \node[symmetrization] (naive-symmetrization) at (0, 0) {Naive};
    \node[symmetrization] (bib-symmetrization) at (0, -1) {Biblio-\\graphical};
  \end{scope}

  \path
    (data-weight.east)
      edge (naive-symmetrization.west)
      edge (bib-symmetrization.west)
    (data-rec-weight.east)
      edge (naive-symmetrization.west)
      edge (bib-symmetrization.west)
    (unit-weight.east)
      edge (naive-symmetrization.west)
      edge (bib-symmetrization.west);

  \begin{scope}[shift={(1.75,.5)}]
    \node[clustering] (bet-clustering) at (0, 0) {Edge\\Betweenness};
    \node[clustering] (agg-clustering) at (0, -1) {Agglomerative};
    \node[clustering] (spec-clustering) at (0, -2) {Spectral};
  \end{scope}

  \path
    (data-weight.east) edge (bet-clustering.west)
    (data-rec-weight.east) edge[bend left=30] (bet-clustering.west)
    (unit-weight.east) edge[bend left=25] (bet-clustering.west)
    (naive-symmetrization.east)
      edge (bet-clustering.west)
      edge (agg-clustering.west)
      edge (spec-clustering.west)
    (bib-symmetrization.east)
      edge (bet-clustering.west)
      edge (agg-clustering.west)
      edge (spec-clustering.west);

  \begin{scope}[shift={(3,1)}]
    \node[criterion] (density-criterion) at (0, 0) {Cluster Density};
    \node[criterion] (local-criterion) at (0, -1) {Local\\Clust. Coeff.};
    \node[criterion] (global-criterion) at (0, -2) {Global\\Clust. Coeff.};
    \node[criterion] (modularity-criterion) at (0, -3) {Modularity};
  \end{scope}

  \path
    (bet-clustering.east)
      edge (density-criterion.west)
      edge (local-criterion.west)
      edge (global-criterion.west)
      edge (modularity-criterion.west)
    (agg-clustering.east)
      edge (density-criterion.west)
      edge (local-criterion.west)
      edge (global-criterion.west)
      edge (modularity-criterion.west)
    (spec-clustering.east)
      edge (density-criterion.west)
      edge (local-criterion.west)
      edge (global-criterion.west)
      edge (modularity-criterion.west);

  \begin{scope}[yshift=2cm]
    \node (weight-label) at (0,0) {Weighting\\Function};
    \node (symmetrization-label) at (.875,0) {Symmetri-\\zation};
    \node (clustering-label) at (1.75,0) {Clustering\\Algorithm};
    \node (stopping-label) at (3,0) {Stopping\\Criterion};
  \end{scope}

  \draw[black!40] (-.5,1.4) -- (3.5, 1.4);

\end{tikzpicture}

%% file: images/MDO_CO_Sellar_labeled.tex
\begin{tikzpicture}[thick,every edge/.style={draw,->,>=stealth}]
  \node[draw,circle,fill=white] (c1) at (0,0) {};
  \node[draw,circle,fill=white] (c2) at (0,-1) {};

  \node[draw,circle,fill=gray] (r1) at (1,-1) {};
  \node[draw,circle,fill=gray] (r2) at (1,0) {};
  \node[draw,circle,fill=gray] (r3) at (2,0) {};
  \node[draw,circle,fill=gray] (r4) at (2,-1) {};

  \node[draw,circle,fill=black] (l1) at (-1,-1) {};
  \node[draw,circle,fill=black] (l2) at (-1,0) {};
  \node[draw,circle,fill=black] (l3) at (-2,0) {};
  \node[draw,circle,fill=black] (l4) at (-2,-1) {};

  \path
    (c1) edge [bend left=10] (c2)
    (c2) edge [bend left=10] (c1)
    (c2) edge (r1)
    (r1) edge (c1) edge (r2) edge[bend left=10] (r3) edge[bend left=10] (r4)
    (r2) edge (c1)
    (r3) edge[bend left=10] (r1) edge (r4)
    (r4) edge [bend left=10] (r1)
    (c2) edge (l1)
    (l1) edge (c1) edge (l2) edge[bend left=10] (l3) edge[bend left=10] (l4)
    (l2) edge (c1)
    (l3) edge[bend left=10] (l1) edge (l4)
    (l4) edge [bend left=10] (l1);
\end{tikzpicture}

%% file: 2021-rce-clustering-workflow-graphs.bbl
\begin{thebibliography}{10}
\providecommand{\url}[1]{\texttt{#1}}
\providecommand{\urlprefix}{URL }
\providecommand{\doi}[1]{https://doi.org/#1}

\bibitem{ApacheNifi}
{Apache Software Foundation}, {Cloudera}, {Hortonworks}: {Apache Nifi},
  \url{https://nifi.apache.org/}

\bibitem{Boden2019RceIntegrationEnvironment}
Boden, B., Flink, J., Mischke, R., Schaffert, K., Weinert, A., Wohlan, A.,
  Schreiber, A.: Rce: An integration environment for engineering and science.
  SoftwareX  \textbf{15} (2021).
  \doi{https://doi.org/10.1016/j.softx.2021.100759}

\bibitem{Brandes2001FasterAlgorithmBetweenness}
Brandes, U.: A faster algorithm for betweenness centrality. Journal of
  mathematical sociology  \textbf{25}(2),  163--177 (2001).
  \doi{10.1080/0022250x.2001.9990249}

\bibitem{BrownSvenson1988}
Brown, M.G., Svenson, R.A.: {Measuring R\&D Productivity}. Research-Technology
  Management  \textbf{31}(4),  11--15 (1988)

\bibitem{Danon2006EffectSizeHeterogeneity}
Danon, L., Diaz-Guilera, A., Arenas, A.: The effect of size heterogeneity on
  community identification in complex networks. Journal of Statistical
  Mechanics: Theory and Experiment  \textbf{2006}(11),  P11010 (2006)

\bibitem{Donetti2004DetectingNetworkCommunities}
Donetti, L., Munoz, M.A.: Detecting network communities: a new systematic and
  efficient algorithm. Journal of Statistical Mechanics: Theory and Experiment
  \textbf{2004}(10),  P10012 (2004). \doi{10.1088/1742-5468/2004/10/p10012}

\bibitem{Du2007AlgorithmDetectingCommunity}
Du, H., Feldman, M.W., Li, S., Jin, X.: An algorithm for detecting community
  structure of social networks based on prior knowledge and modularity.
  Complexity  \textbf{12}(3),  53--60 (2007). \doi{10.1002/cplx.20166}

\bibitem{Dunn2005UseEdgeBetweenness}
Dunn, R., Dudbridge, F., Sanderson, C.M.: The use of edge-betweenness
  clustering to investigate biological function in protein interaction
  networks. BMC bioinformatics  \textbf{6}(1), ~39 (2005)

\bibitem{ElRamlyIglinskiStrouliaEtAl2001Modelingsystemuser}
El-Ramly, M., Iglinski, P., Stroulia, E., Sorenson, P., Matichuk, B.: Modeling
  the system-user dialog using interaction traces. In: WCRE 2001. pp. 208--217
  (2001). \doi{10.1109/WCRE.2001.957825}

\bibitem{Fortunato2010CommunityDetectionGraphs}
Fortunato, S.: Community detection in graphs. Physics reports
  \textbf{486}(3-5),  75--174 (2010). \doi{10.1016/j.physrep.2009.11.002}

\bibitem{Jung2006WorkflowClusteringMethod}
Jung, J.Y., Bae, J.: Workflow clustering method based on process similarity.
  In: ICCSA 2006. pp. 379--389. Springer (2006)

\bibitem{Jung2009HierarchicalClusteringBusiness}
Jung, J.Y., Bae, J., Liu, L.: Hierarchical clustering of business process
  models. International Journal of Innovative Computing, Information and
  Control  \textbf{5}(12),  1349--4198 (2009)

\bibitem{Junker2011AnalysisBiologicalNetworks}
Junker, B.H., Schreiber, F.: Analysis of biological networks, vol.~2. John
  Wiley \& Sons (2011)

\bibitem{KawaguchiGargMatsushitaEtAl2006}
Kawaguchi, S., Garg, P.K., Matsushita, M., Inoue, K.: {MUDABlue}: An automatic
  categorization system for open source repositories. Journal of Systems and
  Software  \textbf{79}(7),  939--953 (jul 2006).
  \doi{10.1016/j.jss.2005.06.044}

\bibitem{Kemper2009ValuationNetworkEffects}
Kemper, A.: Valuation of network effects in software markets: A complex
  networks approach. Springer Science \& Business Media (2009)

\bibitem{Knime}
{KNIME AG}: {KNIME}, \url{https://www.knime.com/}

\bibitem{Luce1949MethodMatrixAnalysis}
Luce, R.D., Perry, A.D.: A method of matrix analysis of group structure.
  Psychometrika  \textbf{14}(2),  95--116 (1949). \doi{10.1007/bf02289146}

\bibitem{VonLuxburg2007TutorialSpectralClustering}
von Luxburg, U.: A tutorial on spectral clustering. Statistics and computing
  \textbf{17}(4),  395--416 (2007). \doi{10.1007/s11222-007-9033-z}

\bibitem{McMillanGrechanikPoshyvanyk2012}
McMillan, C., Grechanik, M., Poshyvanyk, D.: Detecting similar software
  applications. In: ICSE 2012. {IEEE} (2012). \doi{10.1109/icse.2012.6227178}

\bibitem{Newman2001StructureScientificCollaboration}
Newman, M.E.J.: The structure of scientific collaboration networks. Proceedings
  of the national academy of sciences  \textbf{98}(2),  404--409 (2001)

\bibitem{Newman2003MixingPatternsCommunity}
Newman, M.E.J., Girvan, M.: Mixing patterns and community structure in
  networks. In: Statistical mechanics of complex networks, pp. 66--87. Springer
  (2003). \doi{10.1007/978-3-540-44943-0\_5}

\bibitem{Newman2004FindingEvaluatingCommunity}
Newman, M.E.J., Girvan, M.: Finding and evaluating community structure in
  networks. Phys. Rev. E  \textbf{69}(2),  026113 (2004).
  \doi{10.1103/physreve.69.026113}

\bibitem{NguyenDiRoccoRubeiEtAl2018}
Nguyen, P.T., Rocco, J.D., Rubei, R., Ruscio, D.D.: {CrossSim}: Exploiting
  mutual relationships to detect similar {OSS} projects. In: SEAA 2018. {IEEE}
  (aug 2018). \doi{10.1109/seaa.2018.00069}

\bibitem{ModelCenter}
{Phoenix Integration}: {Model Center}, \url{https://www.phoenix-int.com/}

\bibitem{Santos2008FirstStudyClustering}
Santos, E., Lins, L., Ahrens, J.P., Freire, J., Silva, C.T.: A first study on
  clustering collections of workflow graphs. In: IPAW 2008. pp. 160--173.
  Springer (2008)

\bibitem{Satuluri2011SymmetrizationsClusteringDirected}
Satuluri, V., Parthasarathy, S.: Symmetrizations for clustering directed
  graphs. In: EDBT 2011. pp. 343--354 (2011). \doi{10.1145/1951365.1951407}

\bibitem{Schaeffer2007GraphClustering}
Schaeffer, S.E.: Graph clustering. Computer science review  \textbf{1}(1),
  27--64 (2007). \doi{10.1016/j.cosrev.2007.05.001}

\bibitem{Schneider2020ClusteringRCEWorkflow}
Schneider, D.: Clustering of {RCE} workflow graphs (2020),
  \url{https://elib.dlr.de/135996/}, {Bachelor Thesis}

\bibitem{Schneider2021}
Schneider, D.: Workflow clustering tool (2021). \doi{10.5281/zenodo.5121720}

\bibitem{Spielman1996SpectralPartitioningWorks}
Spielman, D.A., Teng, S.H.: Spectral partitioning works: Planar graphs and
  finite element meshes. In: FOCS 1996. pp. 96--105. IEEE (1996).
  \doi{10.1109/sfcs.1996.548468}

\bibitem{Tanaka2012WorkflowSchedulingMinimize}
Tanaka, M., Tatebe, O.: Workflow scheduling to minimize data movement using
  multi-constraint graph partitioning. In: CCGRID 2012. pp. 65--72. IEEE (2012)

\bibitem{Tzerpos1999MojoDistanceMetric}
Tzerpos, V., Holt, R.C.: Mojo: a distance metric for software clusterings. In:
  WCRE 1999. pp. 187--193 (1999). \doi{10.1109/WCRE.1999.806959}

\bibitem{Wasserman1994SocialNetworkAnalysis}
Wasserman, S., Faust, K., et~al.: Social network analysis: Methods and
  applications, vol.~8. Cambridge university press (1994)

\bibitem{Watts1998CollectiveDynamics}
Watts, D.J., Strogatz, S.H.: Collective dynamics of ‘small-world’networks.
  nature  \textbf{393}(6684),  440--442 (1998). \doi{10.1515/9781400841356.301}

\bibitem{Yoon2006AlgorithmModularityAnalysis}
Yoon, J., Blumer, A., Lee, K.: An algorithm for modularity analysis of directed
  and weighted biological networks based on edge-betweenness centrality.
  Bioinformatics  \textbf{22}(24),  3106--3108 (2006).
  \doi{10.1093/bioinformatics/btl533}

\bibitem{ZhangLoKochharEtAl2017}
Zhang, Y., Lo, D., Kochhar, P.S., Xia, X., Li, Q., Sun, J.: Detecting similar
  repositories on {GitHub}. In: SANER 2017. {IEEE} (feb 2017).
  \doi{10.1109/saner.2017.7884605}

\bibitem{ZhihuaWen2003OptimalAlgorithmMojoa}
{Zhihua Wen}, {Tzerpos}, V.: An optimal algorithm for mojo distance. In: IWPC
  2003. pp. 227--235 (2003). \doi{10.1109/WPC.2003.1199206}

\end{thebibliography}
